\documentclass[twocolumn,showpacs,pra,superscriptaddress]{revtex4}
\usepackage{dcolumn}
\usepackage{graphicx}

\newcommand{\be}{\begin{equation}}
\newcommand{\ee}{\end{equation}}
\newcommand{\beq}{\begin{eqnarray}}
\newcommand{\eeq}{\end{eqnarray}}

\newcommand{\ket}[1]{\left| #1 \right\rangle}
\newcommand{\erw}[1]{\left\langle #1 \right\rangle}

\begin{document}

\title{Schemes for robust quantum computation with polar molecules: analysis of experimental feasibility}

\author{Elena Kuznetsova}
\affiliation{Department of Physics, University of Connecticut, Storrs, CT 06269, USA}
\affiliation{ITAMP, Harvard-Smithsonian Center for Astrophysics, 
60 Garden Str., Cambridge, MA 02138, USA}
\author{Robin C\^{o}t\'e}
\affiliation{Department of Physics, University of Connecticut, Storrs, CT 06269, USA}
\author{Kate Kirby}
\affiliation{ITAMP, Harvard-Smithsonian Center for Astrophysics, 
60 Garden Str., Cambridge, MA 02138, USA}
\author{Susanne F. Yelin}
\affiliation{Department of Physics, University of Connecticut, Storrs, CT 06269, USA}
\affiliation{ITAMP, Harvard-Smithsonian Center for Astrophysics, 
60 Garden Str., Cambridge, MA 02138, USA}

\date{\today}

\begin{abstract}

We analyse recently proposed physical implementations of a quantum computer based on polar molecules. A set of general requirements for a molecular system 
is presented, which would provide an optimal combination of quantum gate times, coherence times, number of operations, high gate accuracy and experimental 
feasibility. We proceed with a detailed analysis of a scheme utilizing switchable dipole-dipole interactions between polar molecules. Switchable dipole-dipole 
interaction is an efficient tool for realization of two-qubit quantum gates, necessary to construct a universal set of gates. We consider three possible 
realizations of a phase gate using specific molecules, such as CO, NF, alkali dimers and alkaline-earth monohalides. We suggest suitable electronic states and 
ransitions and investigate requirements for the pulses driving them. Finally, we analyse possible sources of decoherence.

\end{abstract}

\maketitle

\section{Introduction}

Information processing relying on quantum mechanical rather than classical systems holds the promise for a dramatic speedup of such operations as factoring 
large numbers, searches of unstructured databases and simulation of the dynamics of quantum mechanical systems \cite{Nielsen}. Communication using optical 
fields with single photons would provide a level of security impossible with classical communication techniques \cite{QuantComm}. Recent years have witnessed 
remarkable advances in both the theoretical and experimental development of quantum computing technologies, including demonstrations of basic building blocks 
necessary for quantum computing and quantum networking. Various approaches have been explored, but the most advanced are based on trapped ions and 
neutral atoms \cite{opt-lattices}-\cite{ions}, cavity QED \cite{CQED}, liquid NMR \cite{NMR}, and solid-state systems \cite{Solid-state}. Recently dipolar 
molecules were proposed as a system with a set of parameters optimal for physical implementation of quantum computing schemes \cite{DeMille}. Polar molecules combine the advantages of neutral atoms and ions (such as long coherence times, rich level structure, strong optical and microwave transitions, well-developed 
techniques of coherent manipulation with optical and microwave pulses) and of quantum dots and superconducting circuits (easy control with electrostatic fields). Polar molecules are thus compatible with various architectures, including optical lattices, microwave and electrostatic traps and solid-state systems. 
An important aspect of these systems is the ability to access electronic states which exibit a large permanent dipole moment. The molecules can then be 
used for fast conditional dipole-dipole interactions resulting in two-qubit operations, necessary for construction of a universal set of qubit 
gates.   

A number of specific implementations of quantum computing schemes with polar molecules have been suggested. In the original proposal \cite{DeMille} 
the two projections of a permanent dipole moment of a polar molecule on the direction of an external electric field are used to store a qubit. 
Molecules are held in a 1D trap array with a gradient of the electric field, producing a Stark shift specific to each molecule, thus allowing 
molecules to be addressed individually by spectroscopic means. A two-qubit gate is implemented via electric dipole-dipole interaction of molecules. In Ref.\cite{Ostrovskaya} a qubit is encoded in a bound molecular and a free state of two atoms in an optical lattice site. Free atoms can be transferred into the bound molecular state and back with a Raman pulse. If the bound molecular state has a large permanent dipole moment, then two molecules in neighboring sites will dipole-dipole interact producing a phase shift. The phase shift accumulated by a molecule is conditional on the state of another molecule, thus resulting in the phase gate. Ref.\cite{Kotochigova} further develops the idea of Ref.\cite{DeMille} for polar molecules in optical lattices. Tunable dipole-dipole interaction between two molecules can be realized when a microwave field resonant with the $J=0 \rightarrow J=1$ rotational transition in the ground electronic and vibrational state induces a dipole moment in each molecule.

Based on this, a set of general requirements for a dipolar molecular system is proposed, which would provide an optimal combination of long coherence times, short gate times (resulting in maximal number of operations), small gate errors, and experimental feasibility.

1) Choice of qubit states

To store a qubit long-lived states are required, well isolated from the environment, i.e. weakly perturbed by electric and magnetic fields and various 
interactions (for example, dipole-dipole or spin-spin). Good candidates are hyperfine and rotational states of a ground electronic molecular state having a 
negligible permanent dipole moment.

2) Coupling strengths

Fast one and two-qubit gates require the corresponding interaction strengths to be large. The storage states need long lifetimes. This means that the transition between the qubit states $\ket{0}$ and $\ket{1}$ is forbidden and thus Raman transitions could be used to perform one-qubit gates. An alternative approach is to map the qubit to some states coupled by a one-photon allowed transition. In both cases, the resulting transition coupling has to be strong. Choosing electric dipole-allowed transitions with dipole moments of a fraction of a Debye for one-qubit operations, and sufficiently large Rabi frequencies of the laser fields, one qubit gate times can be as small as femtosecond laser pulse durations.

3) Robust dipole-dipole interactions

Possible errors resulting from dipole-diple interactions include (i) instabilities in inter-molecular distance and alignment, and (ii) instabilities of the 
dipole-moment states themselves. To avoid (i), we need either ``frozen'' molecules or a more robust dipole-dipole coupling mechanism such as dipole blockade 
\cite{dipole-blockade}. Permanent dipole states with life-times long compared to the gate times are necessary against (ii).

4) Cooling and trapping

In order to fulfill the requirements of points (2) and (3), the molecule has to be cooled down to sub-Kelvin temperatures (typical rotational transition 
frequencies $\sim 10$ GHz, giving the condition $T\ll 1$ K to avoid population of higher-energy rotational states). In optical lattices and electrostatic traps 
it is necessary to have the molecules occupy a ground motional state of the trap to minimize decoherence and gate error.

5) Decoherence

Provided that a qubit is encoded in hyperfine or rotational sublevels with typical lifetimes of the order of hours and typical optical lattice or other trap 
decoherence times on the order of seconds, the main decoherence mechanisms are (i) limited lifetimes of the (excited) high-dipole states, (ii) spatial 
dependence of the dipole-dipole interaction, and (iii) finite laser-linewidth and other light-induced decoherence. Points (i) and (ii) are discussed above, and 
we expect decoherence times stemming from finite laser linewidth of the order of ms using phase-stabilized lasers \cite{Phase-stabilized}.

Recently we proposed a model of switchable dipole-dipole interaction between polar molecules \cite{paper-PRA}, allowing one to implement 
a universal two-qubit gate, such as a phase gate. In the proposed scheme, the state used to store a qubit and an additional state to switch the interaction
 have significantly different permanent dipole moments. Ideally, the storage state has a zero moment, 
while the "switch" one has a large dipole moment of several Debye, so that two molecules excited to this state will exibit strong dipole-dipole interaction 
and acquire a $\pi$ phase shift. Provided that the molecules can be excited only from one of the qubit states, for example $\ket{1}$, only the $\ket{11}$ state 
of the two-qubit system flips sign, resulting in a phase gate. In this approach only two selected molecules interact, which drastically simplifies the 
two-qubit gate compared to the case when all molecules interact at the same time. In this work we carry out a detailed study of the scheme considering specific 
molecules as examples, and check if it satisfies the general requirements presented above.

\begin{figure}
\center{
\includegraphics[width=6.5cm]{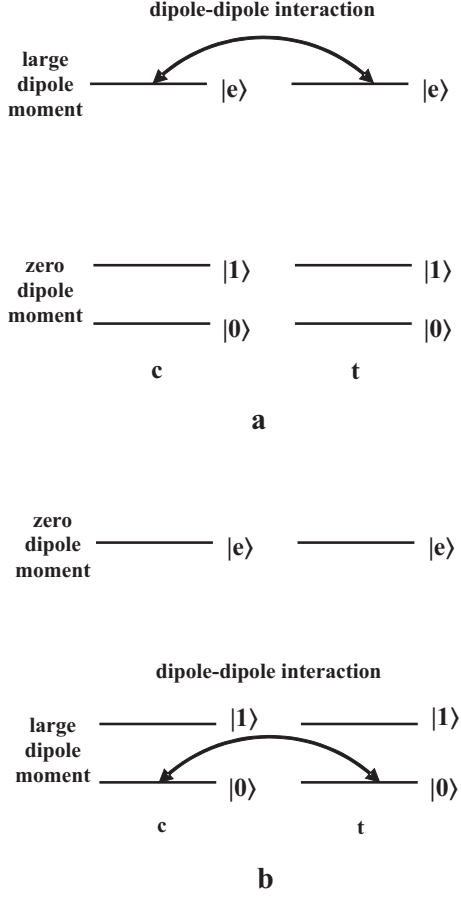}
\caption{\label{fig:gate-schemes}Schematic showing direct (a) and inverted (b) variants of the phase gate based on controllable dipole-dipole 
interaction of polar molecules.}
}
\end{figure}

Three realizations of the phase gate with polar molecules were proposed in previous work. The first one, a direct phase gate, uses a molecular state with a zero permanent dipole moment to store a qubit, while to perform the gate operation, molecules are excited to an additional state with a large dipole moment (see Fig.~(\ref{fig:gate-schemes}a)). We also consider the ``inverted'' case shown in Fig.~(\ref{fig:gate-schemes}b), in which the storage state has a large dipole moment, and the additional state $\ket{e}$ has a small one. In this case the molecules interact and acquire a phase shift when both are in the storage state. The two systems can be implemented using different electronic states of a molecule, as will be shown below. In a ``rotational'' scheme, we could  realize both the direct and inverted schemes using rotational levels of the ground state rather than excited electronic states. The direct and inverted schemes using electronic molecular transitions in the visible or UV range can be most naturally implemented with cold molecules in optical lattices \cite{opt-lattices}. The direct case can also be realized with molecules doped into solid-state matrices. The rotational scheme can be implemented using a recently proposed architecture combining electrostatic traps and coupled microwave resonators \cite{supercond-stripes}.

The organization of the remainder of the paper is as follows. In Section II we analyse the constraints imposed on the relative distance and angles 
of the two interacting molecular dipoles by the requirement to keep a phase error below a threshold value (we set it at 1$\%$). Section III considers 
the direct dipole switching scheme with a CO molecule. The inverted scheme is analysed in Section IV. The scheme based on rotational states is studied 
in Section V. Decoherence mechanisms are analysed in Section VI. Finally, we conclude in Section VII. 

\section{Phase gate error due to molecular movement and dipole misalignment}

\begin{figure}
\center{
\includegraphics[width=7.5cm]{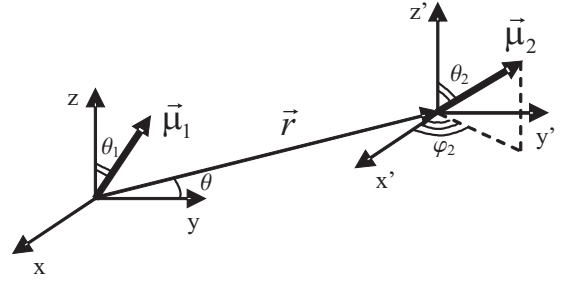}
\caption{\label{fig:angles}Illustration of the calculation of the phase dependence on the relative position and orientation of the 
interacting molecules.}
}
\end{figure}

We analyse the phase errors assuming that two interacting molecules are transferred into states $\ket{e}$ having large permanent dipole moments, 
and the phase is accumulated while they undergo dipole-dipole interaction. The phase dependence on the relative distance and orientation of 
the dipole moments is given by the expression

\begin{eqnarray}
\label{eq:phase}
\phi &\sim& \frac{T}{\hbar}\left( \frac{3(\vec{\mu}_{1}\vec{r})(\vec{\mu}_{2}\vec{r})}{r^{5}}-\frac{\vec{\mu}_{1}\cdot\vec{\mu}_{2}}{r^{3}}\right)= \nonumber \\
&=&\frac{T}{\hbar}\frac{\mu^{2}}{r^{3}}\bigg(3\sin(\theta+\theta_{1})\left(\cos \theta_{2}\sin \theta +\sin \theta_{2} \cos \theta \sin \phi_{2}\right) \nonumber \\
&&\qquad  -\cos \theta_{1}\cos \theta_{2} -\sin \theta_{1} \sin \theta_{2} \sin \phi_{2}\bigg),
\end{eqnarray}

where the angles $\theta $, $\theta_{1}$, $\theta_{2}$ and $\phi_{2}$ determining the orientations of the dipole moments $\vec{\mu}_{1}$ and $\vec{\mu}_{2}$ 
of two interacting molecules are shown in Fig.\ref{fig:angles}, $T$ is the duration of the gate operation, $\vec{r}$ is the vector connecting the first 
dipole with the second one. The vector of the first dipole $\vec{\mu}_{1}$ is assumed to be in the (y,z) plane as well as the $\vec{r}$ vector to simplify 
the analysis (we can always choose the coordinate system this way). The angle $\theta$ describes the vertical shift of the second dipole with respect 
to the first one. The angles $\theta_{1}$ and $\theta_{2}$ are the polar angles of the dipoles in their coordinate systems, $\phi_{2}$ is the azimuthal 
angle of the second dipole. We will analyse the sensitivity of the accumulated phase to the relative distance between the molecules and the angles, 
assuming small deviations from their equilibrium values $r=\erw{r}$, $\theta=\theta_{1}=\theta_{2}=0$, and $\phi_{2}=\pi/2$.

Fidelity of the phase gate, which is directly related to a phase error during the gate operation, will then be affected by the relative motion of the 
molecules, misalignment of their dipole moments, and stability of the duration of optical excitation pulses. From Eq.(\ref{eq:phase}) the phase error 
can be expressed via the uncertainty in the distance $r$ and angles $\theta$, $\theta_{1}$, $\theta_{2}$ and $\phi_{2}$:

\begin{eqnarray}
\label{eq:distance-uncertainty}
\left. \frac{\sqrt{\erw{(\Delta \phi)^{2}}}}{\erw{\phi}}\right|_r &=& 3\frac{\sqrt{\erw{(\Delta r)^{2}}}}{\erw{r}},\\
\label{eq:theta-uncertainty}
\left. \frac{\sqrt{\erw{(\Delta \phi)^{2}}}}{\erw{\phi}}\right|_\theta &=& 3\sqrt{\erw{(\Delta \theta)^{4}}},\\
\label{eq:theta1-uncertainty}
\left. \frac{\sqrt{\erw{(\Delta \phi)^{2}}}}{\erw{\phi}}\right|_{\theta_1} &=& \frac{\sqrt{\erw{(\Delta \theta_{1})^{4}}}}{2},\\
\label{eq:theta2-uncertainty}
\left. \frac{\sqrt{\erw{(\Delta \phi)^{2}}}}{\erw{\phi}}\right|_{\theta_2} &=&\frac{\sqrt{\erw{(\Delta \theta_{2})^{4}}}}{2},\\
\label{eq:phi2-uncertainty}
\left. \frac{\sqrt{\erw{(\Delta \phi)^{2}}}}{\erw{\phi}}\right|_{\phi_2} &=& 0.
\end{eqnarray}

Requiring the phase error be $\le 1\%$ we see from Eq.(\ref{eq:distance-uncertainty}) that the uncertainty of the relative distance has to be $\sqrt{\erw{(\Delta r)^{2}}} \le 0.003\erw{r}$. 
That is for $\erw{r}\sim 500$ nm the molecules have to stay within 1.5 nm of their average distance. As follows from 
Eqs.(\ref{eq:theta-uncertainty})-(\ref{eq:theta2-uncertainty}) the 
same phase error of 1$\%$ results in 
$\sqrt{\erw{(\Delta \theta)^{4}}}\sim 3\cdot 10^{-3}$ (or $\Delta \theta \sim 3^{o}$), and $\sqrt{\erw{(\Delta \theta_{1,2})^{4}}}\sim 2\cdot 10^{-2}$ 
(or $\Delta \theta_{1,2} \sim 8^{o}$).

For a simple estimate of the mean distance a molecule travels in an optical lattice potential minimum we make a harmonic approximation of the 
potential $V_{0}\sin^{2}(kx)\approx V_{0}k^{2}x^{2}$, giving the frequency $\omega=k\sqrt{2V_{0}/m}$ with the corresponding motional ground state wave 
function width $a=\sqrt{\hbar/m\omega}$. Taking $V_{0}=(10-40)E_{R}=(10-40)\hbar^{2}k^{2}/2m$, where $E_{R}=\hbar^{2}k^{2}/2m$ is the recoil energy 
and $(10-40)E_{R}$ is the typical depth of a lattice potential, the width $a$ can be related to the lattice field wavelength $\lambda$ and the 
lattice depth as $a=(10-40)^{-1/4}\lambda/2\pi$. For molecules in neighboring lattice sites $\erw{r}=\lambda/2$, and the phase error according to 
Eq.(\ref{eq:distance-uncertainty}) is $6(10-40)^{-1/4}/2\pi\approx 0.4-0.55$, for five lattice periods distance the phase error is reduced to 
$\sim 10\%$. To reduce the error even further the mean separation between the molecules in the lattice can be increased along with the depth of 
the potential. At the same time this will lead to a decrease in the strength of the dipole-dipole interaction and, therefore, to a larger $T$ 
of the phase gate.

A better solution is to make use of the dipole blockade phenomenon \cite{dipole-blockade}. In this case the excitation scheme B of Ref.\cite{dipole-blockade} 
can be applied, in which the doubly excited state $\ket{ee}$ of the two molecules is shifted due to the dipole-dipole interaction from its unperturbed 
position, the shift being $u \sim \mu^{2}/r^{3}\hbar$, where $\mu$ is the permanent dipole moment of the $\ket{e}$ state. To implement the diplo-blockade 
mechanism the two molecules have to be addressed individually. Then, the first part of the gate operation is the transfer of a control molecule to the excited state $\ket{e}$ with a resonant 
$\pi$ pulse. The second step is to apply to a target molecule a $2\pi$ pulse, resonant with the unperturbed transition $\ket{1}\rightarrow \ket{e}$. Since 
the $\ket{ee}$ state is detuned, the target molecule returns back to $\ket{1}$ acquiring a phase $\tilde{\phi}\sim \pi \Omega_{2\pi}/2u \ll \pi$ 
if $u\gg \Omega_{2\pi}$. In the last step the control molecule is brought back to $\ket{1}$ with a second $\pi$ pulse. The major requirement of the 
dipole-blockade mechanism is to keep the Rabi frequency of the $2\pi$ pulse much smaller than the energy shift of the doubly excited state. As a result 
all states except the $\ket{00}$ one acquire a $\pi$ phase shift, resulting in the phase gate.

Since the exact position of the two molecules is not important if the dipole blockade mechanism is used, it allows one to circumvent the strict 
requirement on the relative distance set by Eq.(\ref{eq:distance-uncertainty}) and on angles by Eqs.(\ref{eq:theta-uncertainty})-(\ref{eq:phi2-uncertainty}). 
The smaller the uncertainties in the relative distance and angles, the better the dipole blockade mechanism works, resulting in smaller fluctuations of $u$. 
The angle $\theta$ can be made small 
by loading molecules in a 3D lattice, which would provide confinement in the vertical direction. Combination of the vertical confinement with the 
dipole-blockade mechanism allows one to satisfy the $\theta$ angle requirement. The $\theta_{1,2}$ angles can be controlled with either static or 
microwave electric fields. Finally, the $\phi_{2}$ dependence as is seen from Eq.(\ref{eq:phi2-uncertainty}) of the phase is weak, provided that the other two angles are close to the optimal values.

The same analysis applies to the architecture utilizing electrostatic traps combined with coupled microwave resonators \cite{supercond-stripes}, where 
the dipole-dipole interaction strength scales as $\mu^{2}/h^{2}r$ with $h$ and $r$ being the distance between the molecule and the trap surface and the 
distance between the traps, respectively. In this case the phase error dependence on the geometry of the trap is given by

\begin{eqnarray}
\label{eq:distance-uncertainty-superc-h}
\frac{\sqrt{\erw{(\Delta \phi)^{2}}}}{\erw{\phi}}&=&2\frac{\sqrt{\erw{(\Delta h)^{2}}}}{\erw{h}},\nonumber \\
\label{eq:distance-uncertainty-superc-r}
\frac{\sqrt{\erw{(\Delta \phi)^{2}}}}{\erw{\phi}} &=& \frac{\sqrt{\erw{(\Delta r)^{2}}}}{\erw{r}} \nonumber.
\end{eqnarray}

For a typical value of $h\sim 1$ $\mu$m the uncertainty of $h$ has to be $\sqrt{\erw{(\Delta h)^{2}}}\le 5$ nm to have the phase error below $1\%$, and 
with typical $\erw{r}\sim 10$ $\mu$m the corresponding uncertainty $\sqrt{\erw{(\Delta r)^{2}}}\le 100$ nm. Again, the requirement for $h$ is hard to 
fulfill unless the dipole blockade mechanism is used.

In the next three sections we will describe three implementations of the phase gate using specific molecules as examples.

\section{Direct phase gate on the basis of CO molecule}

CO molecule has a small permanent dipole moment (0.1 D) in the ground electronic state $X\;^{1}\Sigma^{+}$, which makes it suitable for the direct phase 
gate implementation. We choose as the $\ket{e}$ state the metastable  $a\;^{3}\Pi_{0}$ electronic state with a permanent dipole moment of $1.37$ D 
\cite{Radzig}. If we choose an isotopic variant of CO, the qubit can be encoded in hyperfine sublevels of the ground rovibrational state, weakly coupled 
with other states in the presence of static and non-resonant dynamic electromagnetic fields. We start with an analysis of the hyperfine structure of 
isotopic CO molecules in the $X\;^{1}\Sigma^{+}$ and $a\;^{3}\Pi_{0}$ states.

\subsection{Hyperfine structure of CO}

\begin{figure}
\center{
\includegraphics[width=7.5cm]{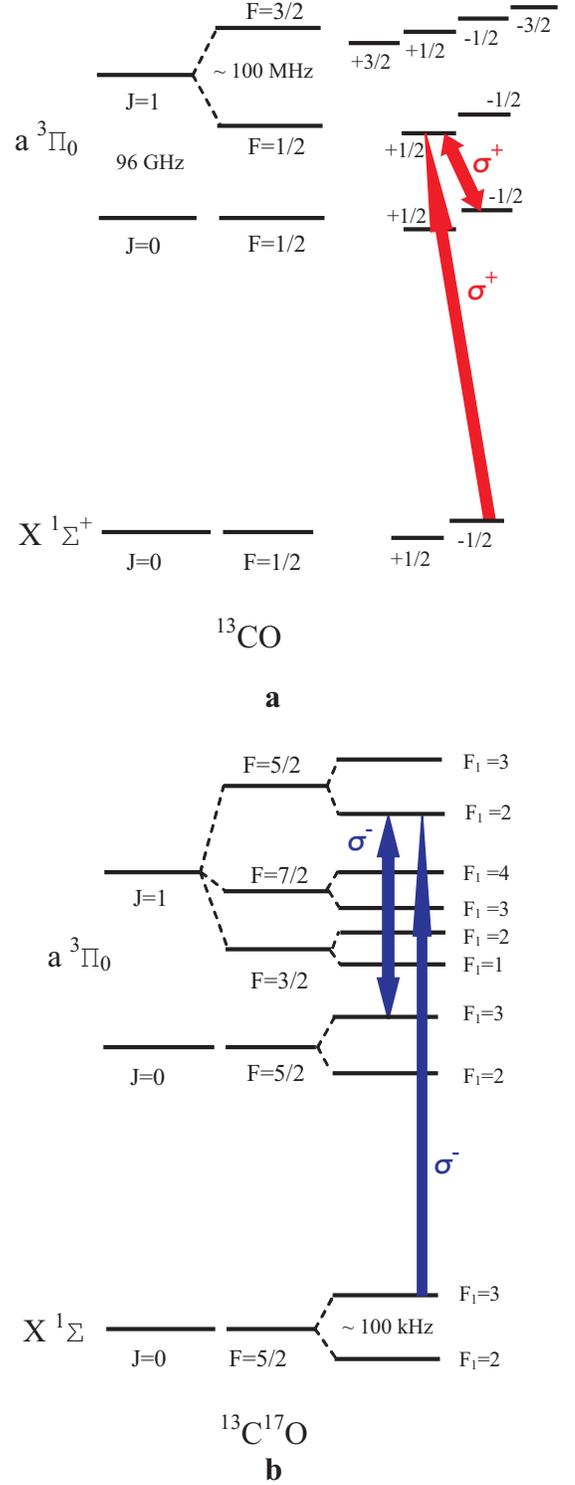}
\caption{\label{fig:Hyperfine-CO} Hyperfine structure and Zeeman splittings of the lowest rotational states of the ground $X\;^{1}\Sigma^{+}$ and excited $a\;^{3}\Pi_{0}$ electronic 
states of two isotopomers of CO along with schemes of selective excitation of $\ket{1}\rightarrow \ket{e}$ transition. 
}}
\end{figure}

There are three CO isotopomers $^{13}$CO ($1\%$ abundance), C$^{17}$O ($0.038\%$ abundance), $^{13}$C$^{17}$O ($3.8\cdot 10^{-4}\%$ abundance) with at least 
one nucleus having a non-zero spin. The $^{13}$C and $^{17}$O nuclear spins are $I_{C}=1/2$ and  $I_{O}=5/2$, respectively. For the ground rotational state 
$J=0$ of $X\;^{1}\Sigma^{+}$ electronic state the coupling with the nuclear spin is described by the Hamiltonian $H_{hfs}=b\vec{J}\vec{I}-eQqF(\vec{I},\vec{J})$, 
where $F(\vec{I},\vec{J})=\left(3C(C+1)/4-I(I+1)J(J+1)\right)/2I(2I-1)(2J-1)(2J+3)$ is the Casimir function (with $C=F(F+1)-I(I+1)-J(J+1)$). For $J=0$ ($F=I$) 
the coupling term vanishes, resulting in zero hyperfine splitting in the ground state for both $^{13}$CO and C$^{17}$O. Small hyperfine splitting is present in 
$^{13}$C$^{17}$O due to interaction of the carbon and oxygen nuclear spins. The level structure of the lowest rotational states of the $X\;^{1}\Sigma^{+}$ and 
$a\;^{3}\Pi_{0}$ states is shown in Fig.\ref{fig:Hyperfine-CO} \cite{CO-hyperfine}. In the case of $^{13}$CO the $\pm 1/2$ nuclear spin sublevels of the 
ground state, Zeeman split by an external magnetic field, can be used as $\ket{0}$, $\ket{1}$ qubit states, respectively. As the $\ket{e}$ state we can 
choose a Zeeman sublevel of $J=1,F=1/2$ hyperfine level of the long-lived $a\;^{3}\Pi_{0}$ ($\nu=0$, $J=1$) state with a lifetime $\sim 0.5$ s. Zeeman 
splitting in the ground electronic state is small, scaling as $\sim 1$ kHz/G, while in the excited state it scales as $\sim 1$ MHz/G due to the non-zero 
electronic angular momentum of the a $^{3}\Pi$ state \cite{Townes}. It means that magnetic fields $\sim 10$ G will suffice to make the Zeeman levels of the 
$\ket{e}$ state resolvable. Selective excitation of $\ket{1}$ can then be realized with a $\sigma^{+}$ laser pulse resonant with the $-1/2 \rightarrow +1/2$ 
transition between the ground and excited electronic states. Single qubit rotations can be performed with Raman pulses using the same $+1/2$ sublevel of 
the $\ket{e}$ state.

In the case of $^{13}$C$^{17}$O the hyperfine sublevels $F_{1}=2,3$ of the ground rovibrational state can be utilized as the $\ket{0}$, $\ket{1}$ states, 
respectively. As the $\ket{e}$ state we can choose the $F_{1}=2$ component of the $J=1,F=5/2$ hyperfine level of the excited  state. Hyperfine splittings of 
$\sim 10-100$ MHz are expected in the $a\;^{3}\Pi$ state due to strong electron spin - nuclear spin interaction, so that the hyperfine structure in this 
state is expected to be resolved with a narrow-band laser. Selective excitation of $\ket{1}$ can be realized then using a $\sigma^{-}$ polarized laser 
pulse resonant with the $J=0,F=5/2,F_{1}=3 \rightarrow J=1,F=5/2,F_{1}=2$ transition, as shown in Fig.\ref{fig:Hyperfine-CO}. Single qubit manipulation 
can also be performed via this excited state hyperfine sublevel using a $\sigma^{-}$ and linearly polarized laser pulses.

Readout of the qubit states can be done in the same manner using the short-lived ($9$ ns lifetime) excited electronic $A\;^{1}\Pi_{1}(J=1,F=1/2)$ state, 
with the corresponding transition wavelength 147 nm. It is the first excited singlet state and therefore it decays directly to the $X\;^{1}\Sigma^{+}$ state. 
Initialization of the qubit into the $\ket{0}$ state can be done by optical pumping using a CW $\sigma^{+}$ (in the case of $^{13}$CO) or $\sigma^{-}$ 
(in the case of $^{13}$C$^{17}$O) polarized laser beam applied in the same way as in Fig.\ref{fig:Hyperfine-CO}.

\subsection{Phase gate operation time}

The spin-forbidden $X\;^{1}\Sigma^{+}\rightarrow $ $a\;^{3}\Pi$ transition is weakly allowed due to the mixing of the $A\;^{1}\Pi_{1}$  and $a\;^{3}\Pi_{1}$ 
states by spin-orbit interaction and rotational mixing between $a\;^{3}\Pi_{1}$ and $a\;^{3}\Pi_{0}$ states \cite{CO-spin-orbit-mixing}. The lifetime of the 
$J=1$ rotational level of the $a\;^{3}\Pi_{0}$ state is $0.5$ s \cite{CO-lifetime}, giving the effective transition dipole moment 
$\mu_{ind} \sim 2\cdot 10^{-4}$ D. If we require that the $\pi$ pulse, transferring population to the $a\;^{3}\Pi$ state has duration 
$T_{\pi}=\pi/\Omega \sim 50$ $\mu$s, the corresponding Rabi frequency and electric field amplitude of the pulse are $\Omega_{\pi}=6\cdot 10^{4}$ s$^{-1}$ and 
$E\sim 10^{2}$ V/cm, with the intensity of the laser pulse $I=cE^{2}/4\pi\sim 25$ W/cm$^{2}$.

As was shown in the previous section the dipole blockade mechanism has to be used to keep the phase error small. For two molecules in neighboring optical 
lattice sites with $r=\lambda/2=100$ nm (assuming that molecules are trapped by a blue-detuned lattice field near resonant with the 
$X\;^{1}\Sigma^{+} - a\;^{3}\Pi$ transition) the shift of the $\ket{ee}$ state due to dipole-dipole interaction is $u=1.87 \cdot 10^{6}$ s$^{-1}$ 
($\approx 300$ kHz), so the gate can be performed with two $\pi$ pulses $\pi_{c}$ exciting and de-exciting a control molecule, and a $2\pi$ pulse $2\pi_{t}$ 
applied to a target molecule. Choosing a Rabi frequency  $\Omega\sim 10^{5}$ s$^{-1}$ for both the $\pi$ and $2\pi$ pulses (satisfying $\Omega\ll u$) 
results in the gate time $T_{gate}=2\pi/\Omega_{\pi}+2\pi/\Omega_{2\pi}\approx 126$ $\mu$s. If the molecules are separated by five lattice periods, for example, 
the energy shift of the $\ket{ee}$ state is $u\approx 1.5\cdot 10^{4}$ s$^{-1}$ ($\approx 2.4$ kHz), the Rabi frequency has to be reduced to fulfill 
$u\gg\Omega$. If we choose $\Omega\sim 2\cdot10^{3}$ s$^{-1}$, the gate time is $T_{gate}=2\pi/\Omega_{\pi}+2\pi/\Omega_{2\pi}\approx 6.3$ ms. Since the 
dipole-blockade mechanism is used, the uncertainty in the positions of the molecules is not affecting the gate.  The resulting phase gate has the following form:

\begin{equation}
\label{eq:oo-CO}
\begin{array}{c@{\quad\stackrel{\pi_c}{\rightarrow}\quad}c@{\quad\stackrel{2\pi_t}{\rightarrow}\quad}c@{\quad\stackrel{\pi_c}{\rightarrow}\quad}c}
\ket{00} & \ket{00} & \ket{00} & \ket{00}, \nonumber \\
\ket{01} & i\ket{0e} & i\ket{0e} & -\ket{01}, \nonumber \\
\ket{10} & \ket{10} & -\ket{10} & -\ket{10}, \nonumber \\
\ket{11} & i\ket{1e} & i\ket{1e} & -\ket{11} \nonumber.
\end{array}
\end{equation}

Another candidate for the direct phase gate is the NF molecule, which has a  small dipole moment $\mu \approx 0.075$ D in the ground electronic 
$X\;^{3}\Sigma^{-}$ state and $\mu \approx 0.75$ D in the excited metastable $b\;^{1}\Sigma^{+}$ state \cite{NF}.

\section{Inverted phase gate with $LiCs$}

The inverted phase gate can be implemented in mixed alkali dimers, many of which have a large dipole moment in their ground electronic state 
$X\;^{1}\Sigma^{+}$ and a small dipole moment in their metastable $a\;^{3}\Sigma^{+}$ state. Heteronuclear alkali dimers are of growing 
interest due to studies of cold collision dynamics and photoassociation processes of laser cooled alkali atoms. We analyze the inverted gate using LiCs, 
which has recently been created experimentally \cite{LiCs-formation} from the ultracold atomic gases Li and Cs.

\subsection{Hyperfine structure of $LiCs$}

\begin{figure}
\center{
\includegraphics[width=7.5cm]{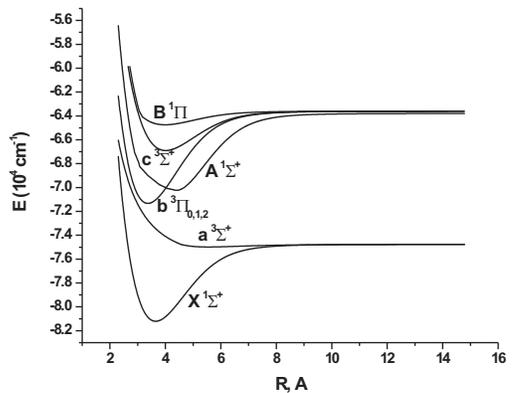}
\caption{\label{fig:LiCs-el-str}Schematic of the low-lying electronic states of LiCs.}
}
\end{figure}

The electronic structure of LiCs was calculated in \cite{LiCs-el-structure}, and a schematic figure of the six lowest electronic states is approximately 
drawn in Fig.\ref{fig:LiCs-el-str}. The metastable $a\;^{3}\Sigma^{+}$ state supports a number of bound states (up to 20 vibrational states were 
observed in NaCs \cite{NaCs}). The ground $X\;^{1}\Sigma^{+}$ state of LiCs has a dipole moment of $\approx 5.5$ D and the metastable state 
has a permanent dipole moment of $\approx -0.45$ D \cite{LiCs-dipole-moments}.

\begin{figure}
\center{
\includegraphics[width=7.5cm]{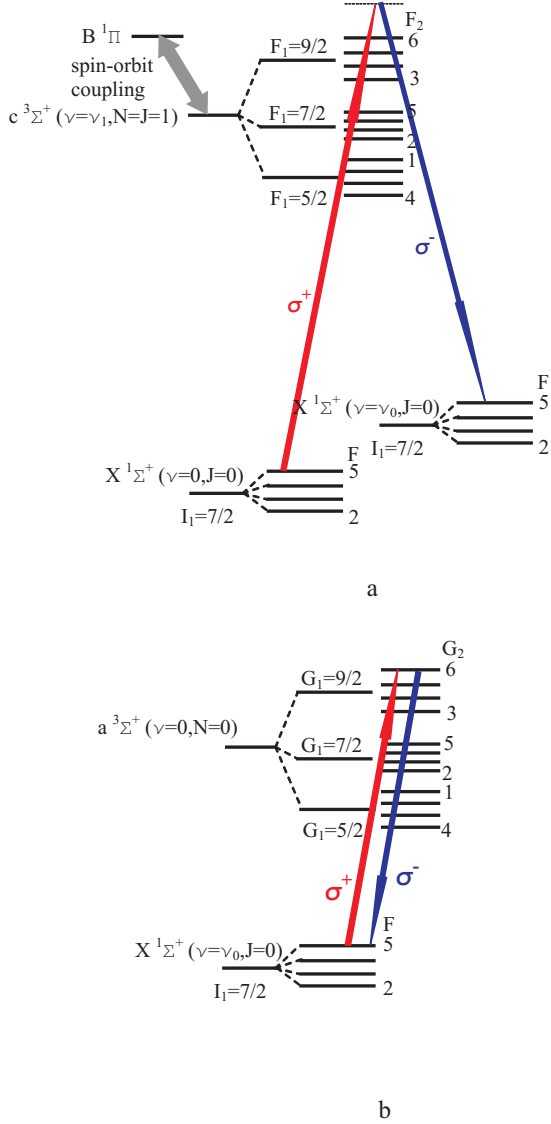}
\caption{\label{fig:Hyperfine-LiCs}Hyperfine structure in the ground $X\;^{1}\Sigma^{+}(v=0,J=0)$, $X\;^{1}\Sigma^{+}(v=v_{0},J=0)$ and excited $a\;^{3}\Sigma^{+}(v=0,N=0,J=1)$, 
$c\;^{3}\Sigma^{+}(v=v_{1},N=J=1)$ states. Hyperfine coupling for $c\;^{3}\Sigma^{+}$ and 
$a\;^{3}\Sigma^{+}$ was found to be Hund's b$_{\beta J}$ and b$_{\beta S}$, respectively \cite{NaRb-hyperf1}. a) Population transfer from $\ket{1}$ state 
($v=0,J=0,F=5$) to a higher vibrational level $\ket{1}'$ ($v=v_{0},J=0,F=5$) of the ground state, having optimal Franck-Condon factor with a ground 
rovibrational state of $a\;^{3}\Sigma^{+}$; b) Selective excitation of $\ket{1}'$ into $\ket{e}$, encoded into $a\;^{3}\Sigma^{+}(v=0,N=0,J=1,G_{1}=9/2,G_{2}=6)$.}
}
\end{figure}

Taking into account that the $^{7}$Li has a nuclear spin $I_{Li}=3/2$, and $^{133}$Cs has a nuclear spin $I_{Cs}=7/2$ the projected ground state 
$X\;^{1}\Sigma$($v=0$,$J=0$) hyperfine structure along with the experimentally observed metastable state $a\;^{3}\Sigma^{+}$ hyperfine structure, 
are shown schematically in Fig.\ref{fig:Hyperfine-LiCs}. The ground state hyperfine structure has not yet been observed in mixed alkali molecules, 
and the splittings are most probably in the kHz - hundreds kHz range. The hyperfine structure of the $c\;^{3}\Sigma^{+}$ and $a\;^{3}\Sigma^{+}$ states 
was observed in $^{23}$Na$^{85}$Rb and in $^{23}$Na$^{39}$K with typical splittings of F$_{2}$ and G$_{2}$ levels $\sim 30-40$ MHz and $\sim 300$ MHz 
\cite{NaRb-hyperf,NaRb-hyperf1,NaK-hyperf}, respectively. For LiCs we assume splittings of the same order. Hyperfine sublevels $F=4$ and $F=5$ of the 
ground rovibrational state can be chosen to encode the qubit $\ket{0}$ and $\ket{1}$ states, respectively. The excited $\ket{e}$ state can be encoded 
in the $G_{1}=9/2,G_{2}=6$ hyperfine sublevel of the metastable $a\;^{3}\Sigma^{+}(v=0,N=0,J=1)$ state. Using circularly polarized laser pulses the 
$\ket{1}$ state can be selectively excited into $\ket{e}$ as shown in Fig.\ref{fig:Hyperfine-LiCs}b, since the $G_{2}$ sublevels are well resolved.

The spin-forbidden $X\;^{1}\Sigma^{+}\rightarrow$ $a\:^{3}\Sigma^{+}$ transition is weakly electric-dipole allowed due to spin-orbit mixing of the 
$a\;^{3}\Sigma^{+}$ state with high energy singlet electronic states. In order to obtain an estimate of an effective dipole moment of the transition we can use 
as examples the known parameters of spin-forbidden transitions in other molecules. Weak fluorescence of the $a\;^{3}\Sigma$ - $X\;^{1}\Sigma^{+}$ was 
observed in BiN, BiP and BiAs \cite{BiN} with radiative lifetimes $\sim 0.5$ ms.  The CO molecule, considered in the previous section, has the 
spin-forbidden $X\;^{1}\Sigma^{+}$ - $a\;^{3}\Pi$ transition with lifetimes ms - hundreds ms depending on the fine and rotational level. If we assume 
the same lifetime of a $^{3}\Sigma^{+}$ for LiCs as for BiN, and take into account the wavelength of the transition $\sim 2$ $\mu$m, the 
corresponding effective transition electric dipole moment is $\mu_{ind}\sim 0.2$ D.

\subsection{Phase gate operation time}

Cold alkali dimers are currently produced from cold alkali atoms using photoassociation and Feshbach resonance techniques 
\cite{RbCs-DeMille,Raman-deexc,Feshbach}. Optical lattices with one molecule per site have been demonstrated \cite{Rb-opt-latt}. Thus alkali dimers are natural 
candidates for quantum computing in optical lattices. As was shown above the phase accumulated during the phase gate strongly depends on the positions of the 
interacting molecules, and it is necessary to utilize the dipole blockade mechanism. The inverted gate can be slightly modified to use it. Let us assume that 
the molecules can be addressed individually. Then in the first step we excite the control molecule from $X\;^{1}\Sigma^{+}(v=0,J=0,F=5)$ ($\ket{1}$ state) to 
some higher-energy vibrational state $X\;^{1}\Sigma^{+}(v=v_{0},J=0,F=5)$ ($\ket{1'}$ state), having significant overlap with the ground vibrational state 
of the $a\;^{3}\Sigma^{+}(v=0,N=0,J=1)$ ($\ket{e}$ state). It can be realized using a STIRAP technique in the way shown in Fig.\ref{fig:Hyperfine-LiCs}a. A $\pi$ 
pulse next transfers the control molecule from $\ket{1'}$ to the lowest rovibrational state of the $a\;^{3}\Sigma^{+}$ state ($\ket{e}$ state). 

Next the target molecule is Raman-excited into $\ket{1'}$, and a $2\pi$ pulse resonant to the transition $\ket{1'}\rightarrow \ket{e}$ is applied. 
The control molecule is then deexcited with a second $\pi$ pulse into $\ket{1'}$, and finally both molecules are Raman transferred back into $\ket{1}$. 
If the control molecule is in the $\ket{0}$ state, it is not excited to $\ket{e}$, and interacts strongly with the target molecule, shifting the 
$\ket{1'}\rightarrow \ket{e}$ transition frequency. The probability of excitation of the target molecule to the $\ket{e}$ state during the $2\pi$ is therefore 
very small. 
It will result in a negligible phase shift, significantly less than $\pi$, accumulated during the $2\pi$ pulse, 
thus implementing the phase gate.

The total gate operation time comes from the STIRAP sequence and the $\pi$ pulse for the control molecule, followed by the STIRAP sequence and the $2\pi$ 
pulse for the target molecule and the STIRAP deexcitation. Final contribution is from the control molecule $\pi$ pulse and STIRAP deexcitation sequence. 
The intermediate $c\;^{3}\Sigma^{+}$ state lifetime is $\sim 20-27$ ns, so the STIRAP pulses have to be of 100 ps-1 ns duration. The duration of the 
$\pi$ pulse between the $X\;^{1}\Sigma^{+}(v_{0},J=0)$ and $a\;^{3}\Sigma^{+}(v=0,N=0,J=1)$ states can be set at $T_{\pi}=50$ $\mu$s as for the CO case, 
with the $2\pi$ pulse duration of 100 $\mu$s, respectively. For the 50 $\mu$s $\pi$ pulse the required Rabi frequency is $\Omega_{\pi}=6\cdot 10^{4}$ s$^{-1}$, 
resulting in the amplitude of the electric component of the laser pulse $E\sim 0.15$ V/cm, with a corresponding laser pulse intensity 
$I\sim 70$ $\mu$W/cm$^{2}$. The $2\pi$ pulse Rabi frequency has to satisfy two conditions: first, $\Omega_{2\pi} \ll u$ to be able to use the dipole blockade 
mechanism due to the interaction of the molecules in the ground electronic state; second, it should take into account that the dipole moment of the 
$\ket{e}$ state in the case of LiCs is not really zero. The Rabi frequency has to exceed the shift of the $\ket{e1}$ state due to dipole-dipole interaction 
between the control molecule in $\ket{e}$ and the target molecule in the $\ket{1}$ state to have the $2\pi$ pulse resonant. Taking the $\ket{e}$ state 
dipole moment $\approx 0.45$ D, and assuming that a blue-detuned lattice laser field for LiCs will have a wavelength $\lambda \sim 800$ nm, 
for two molecules in neighboring sites the energy shift of the $\ket{1}\rightarrow \ket{e}$ transition due to dipole-dipole interaction 
is $u_{e1}\sim 4\cdot 10^{4}$ s$^{-1}$ and the 2$\pi$ Rabi frequency of $\Omega_{2\pi}=10^{5}$ s$^{-1}$ will suffice. On the other hand the 
energy shift of the $\ket{1}\rightarrow \ket{e}$ transition due to dipole-dipole interaction of the molecules in the ground electronic state 
is $u\sim 5\cdot 10^{5}$ s$^{-1}$ and $u\gg \Omega_{2\pi}$, allowing one to use the dipole-blockade mechanism. The STIRAP pulses are much shorter 
than the $\pi$ and $2\pi$ pulses, so their contribution to the phase gate is small, and the total phase gate operation time is 
$T_{gate}=2\pi/\Omega_{\pi}+2\pi/\Omega_{2\pi}\sim 160$ $\mu$s.

From the rotational constant of the ground electronic state of LiCs $B=0.1935$ cm$^{-1}$ and its dipole moment $\mu$ \cite{LiCs-el-structure} the magnitude 
of a static electric field necessary to efficiently mix the $J=0$ and $J=1$ rotational states of the ground state can be calculated 
$\cal E$ $\sim 2B\hbar/\mu\approx 4$ kV/cm. The electric field is switched on only during the $2\pi$ pulse.

The gate we obtain after the sequence of pulses described above is:

\begin{equation}
\label{eq:oo}
\begin{array}{c@{\quad\stackrel{\pi_c}{\rightarrow}\quad}c@{\quad\stackrel{2\pi_t}{\rightarrow}\quad}c@{\quad\stackrel{\pi_c}{\rightarrow}\quad}c}
\ket{00} & \ket{00} & e^{i(\Phi_{c}+\Phi_{t})}\ket{00} & e^{i(\Phi_{c}+\Phi_{t})}\ket{00} ,\\
\ket{01} & i\ket{0e} & ie^{i\tilde{\Phi_{t}}}\ket{0e} & -e^{i\tilde{\Phi_{t}}}\ket{01} ,\\
\ket{10} & \ket{10} & e^{i(\Phi_{c}+\Phi_{t})}\ket{10} & e^{i(\Phi_{c}+\Phi_{t})}\ket{10} ,\\
\ket{11} & i\ket{1e} & -ie^{i\tilde{\Phi_{t}}/2}\ket{1e} & e^{i\tilde{\Phi_{t}}/2}\ket{11} ,
\end{array}
\end{equation}

where in the third line of Eq.(\ref{eq:oo}) we made use of the detuning of the state $\ket{10}$ from its unperturbed position due to the dipole-dipole interaction, so that 
the $2\pi$ pulse applied to the target molecule is not resonant and does not change its phase. In the last line of Eq.(\ref{eq:oo}) line we took into account that during a 
resonant $2\pi$ pulse the target molecule spends half of the pulse duration in $\ket{1}$, and therefore, accumulates a phase shift $\tilde{\Phi_{t}}/2$. 
The phase shifts $\Phi_{t,c}$ and $\tilde{\Phi_{t}}$ are accumulated during the $2\pi$ pulse due to dipole-dipole interaction with all molecules in the 
lattice and due to evolution in the DC electric field:

\begin{eqnarray}
\label{eq:Phi_ct}
\Phi_{c,t} &=& \sum_{j\ne c,t}\frac{\mu^{2}T}{r_{c,t\;j}^{3}\hbar}+\mu {\cal E} T_{2\pi}/\hbar,\\
\label{eq:Phi_t}
\tilde\Phi_{t} &=&\sum_{j\ne c,t}\frac{\mu^{2}T}{r_{c,t\;j}^{3}\hbar}-\frac{\mu^{2}T}{r_{c,t}^{3}\hbar}+\mu {\cal{E}} T/\hbar+\mu_{e} {\cal E} T_{2\pi}/\hbar,
\end{eqnarray}

where $\mu$ and $\mu_{e}$ are the $X\;^{1}\Sigma^{+}$ and $a\;^{3}\Sigma^{+}$ states permanent electric dipole moments.

Choosing ${\cal E}$ sufficiently large so that the electric field induced phase is much greater than the interaction phase, we can set 
$\Phi_{c,t}\approx \tilde{\Phi}_{t}=\pi(2n+1)$, resulting in the gate:

\begin{eqnarray}
\label{00-inv}
\ket{00} &\rightarrow& \ket{00}, \nonumber \\
\label{01-inv}
\ket{01} &\rightarrow& \ket{01}, \nonumber \\
\label{10-inv}
\ket{10} &\rightarrow& \ket{10}, \nonumber \\
\label{11-inv}
\ket{11} &\rightarrow& i\ket{11} \nonumber .
\end{eqnarray}

The resulting gate gives only "half" of the true phase gate, so we need to perform the operation two times to obtain the phase gate, in which 
$\ket{11}$ changes sign.

Electric fields required to mix the $J=0$ and $J=1$ rotational states of the ground state in order for the $J=0$ state to acquire a dipole moment will 
produce an interaction term $\mu {\cal E}/\hbar \sim 7\cdot 10^{10}$ s$^{-1}$, exceeding the dipole-dipole interaction strength $u$ between two neighboring 
molecules, calculated above, by five orders of magnitude. It means that the electrostatic interaction contribution to the phase given in Eqs.(\ref{eq:Phi_ct}), 
(\ref{eq:Phi_t}) would exceed the dipole-dipole interaction contribution from more than $10^{5}$ molecules in the lattice, and 
$\Phi_{c,t},\;\tilde{\Phi_{t}}\approx \mu {\cal E}T_{2\pi}/\hbar$. In order to keep the phase error below $\%1$ level, the electric field magnitude 
will have to be controlled with better than $5$ $\mu$V/cm precision for $T_{2\pi}=100$ $\mu$s.

\section{Rotational scheme based on $BaI$}

To implement the rotational scheme we can use alkaline earth monohalides each having $X\;^{2}\Sigma^{+}$ ground state with a large dipole moment and 
significant hyperfine splittings. Specifically we consider $^{138}$Ba$^{127}$I, $^{138}$Ba$^{79}$Br and $^{88}$Sr$^{127}$I, $^{88}$Sr$^{79}$Br with 
permanent ground state dipole moments $\mu \approx 5.5-6$ D \cite{alkali-halides}. 

\begin{figure}
\center{
\includegraphics[width=6.5cm]{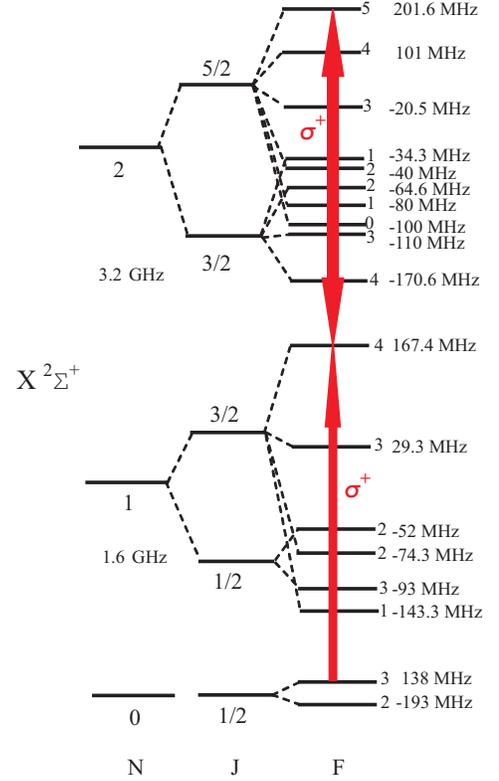}
\caption{\label{fig:Hyperfine-BaI}Proposed scheme of selective excitation of $|1>$ state (encoded in the F=3 hyperfine sublevel of the N=0 rotational 
level of the ground vibrational state of $X\;^{2}\Sigma^{+}$ electronic level of BaI 
with $|0>$ as 
F=2 state) into $|e>$ state (a hyperfine sublevel F=4 of the N=1 rotational state) using a microwave $\sigma^{+}$ pulse. The second microwave pulse, 
coupling the $J=1$ and $J=2$ states serves as 
a dressing field, preventing from non-radiative energy exchange between two interacting as well as 
other molecules, which are in the ground $N=0$ state. 
}
}
\end{figure}

\subsection{Hyperfine structure of $BaI$}

Rotational, fine and hyperfine structure of the ground $X\;^{2}\Sigma^{+}$ state of $^{138}Ba^{127}I$ can be calculated from the Hamiltonian

\begin{equation}
\hat{H}=\hat{H}_{rot}+\hat{H}_{sr}+\hat{H}_{hfs},
\end{equation}

including the rotational $\hat{H}_{rot}=B_{e}\hat{\vec{N}}^{2}$, spin-rotational $\hat{H}_{sr}=\gamma_{sr}\hat{\vec{N}}\hat{\vec{S}}$ and hyperfine 
$\hat{H}_{hfs}=b_{F}\vec{\hat{I}}\vec{\hat{S}}+cI_{z'}S_{z'}-eQq\left[3\hat{I}^{2}_{z}-I(I+1)\right]/4I(I-1)$ terms \cite{BaI-2Sigma-Hamilt}. The 
coupling constants are $\gamma_{sr}=75.85$ MHz, $b_{F}=93.117$ MHz, $c=52.17$ MHz and $eQq=33.62$ MHz \cite{BaI-2Sigma-Hamilt}. The spin-rotation and 
hyperfine terms are comparable, and the coupling of rotational, electron and nuclear spin angular momenta in the lowest rotational states 
is intermediate between the Hund's $b_{\beta J}$ and $b_{\beta S}$ cases \cite{SrI-hyperfine}. The energies of the hyperfine levels were calculated 
using $b_{\beta J}$ coupling case \cite{Townes}. The hyperfine structure is due to the nuclear spin of $^{127}$I I=5/2 (100$\%$ abundance), 
since the most abundant $^{138}$Ba isotope (71$\%$) has zero nuclear spin. The scheme of the lowest rotational states of BaI including hyperfine structure 
is given in Fig.\ref{fig:Hyperfine-BaI}.

\subsection{Phase gate operation time}

\begin{figure}
\center{
\includegraphics[width=7.5cm]{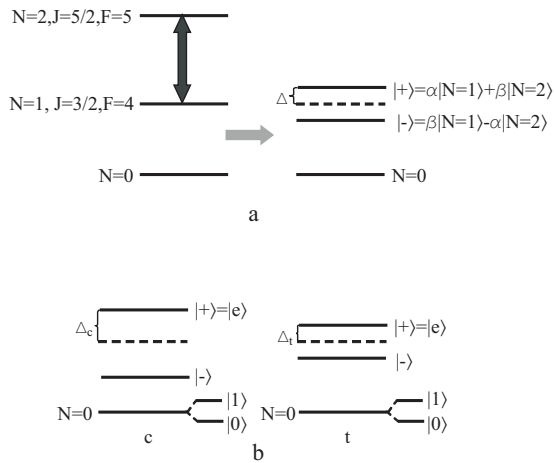}
\caption{\label{fig:flip-flop-suppression}a) Proposed scheme of suppression of non-radiative energy exchange between a pair of molecules; 
b) Hyperfine sublevels of the ground rotational state $N=0$ used to encode qubits in the control and target molecules. 
}
}
\end{figure}

The qubit $\ket{0}$ and $\ket{1}$ states can be encoded into the $F=2$ and $F=3$ hyperfine sublevels of the ground rovibrational state. The state $\ket{e}$ 
can be a hyperfine sublevel of either the $N=1$ or $N=2$ rotational state. On the one hand, it is preferable to choose the $N=2$ state since there is 
no electric-dipole allowed transition to the ground $N=0$ state, meaning that non-radiative energy exchange processes between pairs of molecules 
due to electric dipole-dipole coupling, one in $N=0$ and the other in $N=2$ states are prohibited. On the other hand, the molecules can be excited to the 
$N=2$ state only via two-photon processes, always resulting in a $\pi$ phase shift of the wavefunction (not the desired $\pi/2$ phase shift), which does 
not allow one to realize the phase gate. The solution can be to use a superposition of $N=1$ and $N=2$ levels, produced by a microwave field resonantly 
coupling these states, as shown in Fig.\ref{fig:flip-flop-suppression}a. 

The resulting dressed states $\ket{+}$ and $\ket{-}$ are shifted from the unperturbed $N=1$ level position by $\Delta_{c,t}=\pm \Omega_{c,t\;coupl}$, 
which is the effective Rabi frequency of coupling fields. The non-radiative energy exchange between two molecules has a strength 
$V_{ind-dip}\sim \mu^{2}/r^{3}\hbar$, where $\mu$ is the dipole moment of the corresponding transition, $r$ is the distance between the molecules. For 
rotational transitions we consider the dipole moment is given by the permanent dipole moment of the ground electronic state. 
If $|\Delta_{c,t}| \gg V_{ind-dip}$ the energy exchange between the target and the control molecule as well as with other molecules in $N=0$ states is 
strongly suppressed, since the two molecules are no longer in resonance. Here we should note that for transitions between rotational levels of the same 
electronic and vibrational state the transition dipole moment is given by the permanent dipole moment of the ground electronic state, so that the natural 
linewidth of a rotational level can be roughly calculated as $A=4\mu^{2}(2\pi)^{3}/3\lambda_{mw}^{3}\hbar$, where $\lambda_{mw}$ is the wavelength of the 
$N\rightarrow N-1$ transition. Since $\lambda_{mw}$ is larger or comparable to a typical distance $r$ between molecules in an optical lattice or in 
electrostatic traps, $A\le V_{ind-dip}$ and the exchange processes are strong. For the case of CO and LiCs analysed in previous sections, $r\ge \lambda/2$ 
meaning that $A\ge V_{ind-dip}$ and the exchange processes are slower than the spontaneous decay and can be neglected. In the case of BaI the 
$\ket{N=1,J=3/2,F=4}$ and $\ket{N=2,J=5/2,F=5}$ states can be mixed to form $\ket{+}$ and $\ket{-}$ states in the control and target qubits. 
Fig.\ref{fig:flip-flop-suppression}b illustrates the way qubits are encoded in hyperfine sublevels of the $N=0$ state aa well as the coupling of the $N=1$ 
and $N=2$ states to form the $\ket{e}$ state. The phase gate in this case is direct since the ground $N=0$ state does not possess a dipole moment. 
The gate can be implemented in the same way as in the CO case. First, a $\pi$ pulse excites the control qubit into $\ket{e}=\ket{+}$ state. Next a $2\pi$ 
pulse, resonant with the unshifted $\ket{1} \rightarrow \ket{+}=\ket{e}$ transition is applied to the target molecule. If the control qubit is in 
the $\ket{e}$ state, the doubly excited $\ket{ee}$ state will be shifted in energy by the $V_{dip}=\mu^{2}/r^{3}\hbar$ and the $2\pi$ pulse will 
be out of resonance, producing a small phase shift of the wavefunction significantly less than $\pi$. Finally, the control molecule is returned back to 
$\ket{1}$ by a second $\pi$ pulse.

As was already discussed above, Raman pulses used to excite a molecule from $\ket{1}$ to $\ket{e}$ always bring about a $\pi$ shift of the wavefunction, 
which does not allow one to realize the phase gate. In turn, it means that the rotational scheme can be implemented only using one-photon microwave 
pulses, directly driving rotational transitions, making this scheme suitable for the architecture involving electrostatic traps and coupled microwave 
resonators \cite{supercond-stripes}. Let us estimate the time required to implement the gate. As was shown the splitting of the dressed states $\ket{+}$ and 
$\ket{-}$ has to exceed the exchange interaction strength $V_{ind-dip}=\mu^{2}/h^{2}r\hbar$ (for coupled resonators architecture). Taking $\mu\approx 6$ D, 
$h=0.1$ $\mu$m, $r=10$ $\mu$m we have $V_{ind-dip}\sim 3.6\cdot 10^{5}$ s$^{-1}$. So we can take $\Omega_{c,coupl}\sim 3\cdot 10^{7}$ s$^{-1}$ and 
$\Omega_{t,coupl}\sim 3\cdot 10^{6}$ s$^{-1}$, corresponding to intensities of the coupling MW fields $I_{c}\sim 6$ mW/cm$^{2}$ and $I_{t}\sim 60$ 
$\mu$W/cm$^{2}$, respectively. The Rabi frequency of the $\pi$ pulses used to manipulate the control molecule has to be less than $\Delta_{c}$ to be able 
to couple the ground state to only one of the dressed states. We can take $\Omega_{\pi}=3\cdot 10^{5}$ s$^{-1}$, resulting in the duration of the pulse 
$T_{\pi}=\pi/\Omega_{\pi}=10$ $\mu$s. The target qubit $2\pi$ pulse Rabi frequency has to be $\Omega_{2\pi}\ll V_{dip}\approx V_{ind-dip}$, so we can take 
$\Omega_{2\pi}=2\cdot 10^{4}$ s$^{-1}$. The resulting duration of the $2\pi$ pulse is $T_{2\pi}=2\pi/\Omega_{2\pi}\approx 300$ $\mu$s. So the total phase gate 
time is $T_{gate}=320$ $\mu$s. This time is significantly smaller than the natural decay time of rotational states of the ground electronic and vibrational 
state of a polar molecule (typical numbers are $10^{5}$ s).

The state of the qubit can be read out using a microwave field tuned far from resonance with the qubit transition \cite{supercond-stripes}. After the field 
interacts with the qubit transition, it accumulates a phase, which depends on the state of the qubit, allowing one to read out the qubit state from the 
transmission or reflection spectra of the field.

The ground rovibrational state $N=0$ of BaI is strong-field seeking, and there are proposals as to how to design an electrostatic trap of this type. For a pair 
of molecules, each trapped above a small conductor, with the conductors connected by a superconducting wire, the dipole-dipole interaction is modified and has 
a form $V_{dip}\sim \mu^{2}/h^{2}r$. Actually, this is strictly true in the case when the temporal dynamics of interacting dipoles is slower than the 
characteristic frequency of the wire, scaling as $\omega=n\pi v/r$, where $n$ is an integer, and $v$ is the transmission velocity of the wire. Taking 
$r\sim 10$ $\mu$m, and $v\sim c$, where $c$ is the speed of light, we have the lowest frequency of the wire $\omega_{min}\approx 10^{13}$ s$^{-1}$. It 
corresponds to characteristic times $\sim 100$ fs, significantly smaller than any processes in our system. It means that the charge distribution in the 
wire will adiabatically follow the dynamics of the dipoles, and the coupling between the dipoles can be considered as static. The mixing between $N=1$ and 
$N=2$ states can be realized in a microwave cavity as was proposed in Ref.\cite{supercond-stripes}, the coupling strength can be varied by tuning the 
frequency of the $N=1\leftrightarrow N=2$ transition in and out of resonance with the cavity mode by the trap DC electric field. Microwave $\pi$ and $2\pi$ 
pulses can be realized using classical microwave fields, resonant to the $\ket{1}\rightarrow \ket{e}$ transition of a selected molecule.

\section{Decoherence mechanisms}

In the proposed schemes, qubits are stored in hyperfine sublevels of a ground rovibrational electronic state, which makes them insensitive (at least in 
first order) to local fluctuations of DC and AC electric fields. The sublevels do feel fluctuations in a magnetic field, which should be minimized on a time scale of 
a second, relevant to a lifetime of a molecule in an optical lattice and an electrostatic trap. Lifetime of a nuclear spin state of a single molecule, 
isolated from the environment, can be as long as years. Lifetimes of ultracold 
molecules in a far-detuned optical lattice of $\approx 1$ s have been obtained by minimizing scattering of lattice photons \cite{lattice-lifetime}.

Since all three schemes rely on the dipole-blockade mechanism, it is not critical (although it is desirable) to cool molecules to the motional ground state of a lattice or an electrostatic 
trap. It is experimentally possible to load BEC atoms, from which molecules can be formed using photoassociation or Feshbach resonance techniques, into the ground 
state of a lattice by switching it on adiabatically \cite{lattice-switching}. In an electrostatic trap sideband cooling using microwave resonator enhanced 
spontaneous emission was proposed for efficient cooling of molecules to the ground state of motion \cite{supercond-stripes}. With overall coherence time of the 
order of 1 s, the maximal number of operations, which can be performed with the direct, inverted and rotational gates, is $\sim 8\cdot 10^{3},\;6\cdot 10^{3}$ and 
$3\cdot 10^{3}$, respectively.

\section{Conclusion}
Analysis of quantum computation proposals with polar molecules allows one to define a set of requirements for a molecular system optimizing its performance 
in terms of quantum gate time, number of operations, coherence time, gate error and experimental simplicity. A scheme studied in detail in the present work 
utilizes switchable dipole-dipole interaction between molecules, which can be used to implement a universal two-qubit gate such as a phase gate. 
It was demonstrated that proposed schemes of the phase gate realization using polar molecules satisfy the set of requirements presented in the introduction, and 
are experimentally 
feasible with currently existing technologies. For the direct phase gate the CO and NF molecules are suitable candidates, and we are looking for other 
molecules with the same parameters, while for the inverted and rotational schemes two classes of molecules are suggested, namely, alkali dimers and 
alkaline-earth monohalides, respectively. Optical lattice and architectures using electrostatic traps and coupled microwave resonators combined 
with the dipole-blockade effect can be used to implement the phase gate with high fidelity. Decoherence present in both architectures 
limits the qubit lifetime to $\sim 1$ s, allowing several thousand operations to be performed.       

\section{Acknowledgements} 
The authors gratefully acknowledge financial support from ARO and NSF.


\begin{thebibliography}{11}

\bibitem{Nielsen} M.A.Nielsen, I.L.Chuang, {\it Quantum Computation and Quantum Information} (Cambridge University Press, Cambridge, 2000).

\bibitem{QuantComm} {\it Quantum Communications and Cryptography}, edited by A.V.Sergienko (Boca Raton, FL: Taylor \& Francis, 2006).

\bibitem{opt-lattices} D.Jaksch, Contempor. Phys. {\bf 45}, 367 (2004).

\bibitem{ions} J.I.Cirac, P.Zoller, Nature (London) {\bf 404}, 579 (2000); D.Kielpinski, C.Monroe, D.J.Wineland, Nature (London) {\bf 417}, 709 (2002).

\bibitem{CQED} A.Rauschenbeutel, G.Nogues, S.Osnaghi, P.Bertet, M.Brune, J.M.Raimond, S.Haroche, Science {\bf 288}, 2024 (2000); J.M.Raimond, M.Brune, S.Haroche, 
Rev. Mod. Phys. {\bf 73}, 565 (2001).

\bibitem{NMR} L.M.K.Vandersypen, I.L.Chuang, Rev. Mod. Phys. {\bf 76}, 1037 (2004).

\bibitem{Solid-state} B.E.Kane, Nature (London) {\bf 393}, 133 (1998); D.Loss, D.P.DiVincenzo, Phys. Rev. A {\bf 57}, 120 (1998); 
Y.Makhlin, G.Schon, A.Shnirman, Rev. Mod. Phys. {\bf 73}, 357 (2001); 

\bibitem{DeMille} D.DeMille, Phys. Rev. Lett. {\bf 88}, 067901 (2002).

\bibitem{Ostrovskaya} C.Lee, E.A.Ostrovskaya, Phys. Rev. A {\bf 72}, 062321 (2005).

\bibitem{Kotochigova} S.Kotochigova, E.Tiesinga, Phys. Rev. A {\bf 73}, 041405(R) (2006).

\bibitem{dipole-blockade} D.Jaksch, J.I.Cirac, P.Zoller, S.L.Robinson, R.C$\hat{o}$t\'e, M.D.Lukin, Phys. Rev. Lett. {\bf 85}, 2208 (2000).

\bibitem{Phase-stabilized} J.J.Longdell, M.J.Sellars, N.B.Manson, Phys. Rev. Lett. {\bf 93}, 130503 (2004).

\bibitem{paper-PRA} S.F.Yelin, K.Kirby, R.C$\hat{o}$t\'e, Phys. Rev. A {\bf 74}, 050301(R) (2006). 

\bibitem{supercond-stripes} A.Andr\'e, D.DeMille, J.M.Doyle, M.D.Lukin, S.E.Maxwell, P.Rabl, R.J.Schoelkopf, P.Zoller, Nat. Phys. {\bf 2}, 636 (2006). 

\bibitem{Radzig} A.A.Radzig, B.M.Smirnov, {\it Reference Data on Atoms, Molecules, and Ions} (Springer-Verlag, Berlin, Heidelberg, New York, Tokyo, 1985).

\bibitem{CO-hyperfine} C.Puzzarini, L.Dore, G.Cazzoli, J. Mol. Spectr. {\bf 217}, 19 (2003).

\bibitem{Townes} C.H.Townes, A.L.Schawlow, {\it Microwave Spectroscopy}, (Dover Publications, Inc., New York, 1975). 

\bibitem{CO-spin-orbit-mixing} T.C.James, J. Chem. Phys. {\bf 55}, 4118 (1971).

\bibitem{CO-lifetime} T.Sykora, C.R.Vidal, J. Chem. Phys. {\bf 110}, 6319 (1999).

\bibitem{Phase-gate} E.Charron, P.Milman, A.Keller, O.Atabek, Phys. Rev. A {\bf 75}, 033414 (2007).


\bibitem{NF} S.Kardahakis, J.Pittner, P.Carsky, A.Mavridis, Int. J. Quant. Chem. {\bf 104}, 458 (2005).

\bibitem{Solids} D.E.Milligan, M.E.Jacox, J. Chem. Phys. {\bf 40}, 2461 (1964); J.Bahrdt, N.Schwentner, Chem. Phys. {\bf 127}, 263 (1988); S.Tam, M.E.Fajardo, 
J. Low Temp. Phys. {\bf 122}, 345 (2001).

\bibitem{Solids-translations} H.Friedmann, S.Kimel, Phys. Rev. {\bf 43}, 3925 (1965); H.K.Shin, J. Chem. Phys. {\bf 75}, 3821 (1981).

\bibitem{Solid-Ar} A.C.Becker, J.Langen, H.M.Oberhoffer, U.Schurath, J. Chem. Phys. {\bf 84}, 2907 (1986).

\bibitem{LiCs-formation} S.D.Kraft, P.Staanum, J.Lange, L.Vogel, R.Wester, M.Weidemuller, J. Phys. B {\bf 39}, S993 (2006).

\bibitem{LiCs-el-structure} M.Korek, A.R.Allouche, K.Fakhreddine, A.Chaalan, Can. J. Phys. {\bf 78}, 977 (2000).

\bibitem{NaCs} O.Docenko, M.Tamanis, J.Zaharova, R.Ferber, A.Pashov, H.Kn\"okel, E.Tiemann, J. Phys. B: At. Mol. Opt. Phys. {\bf 39}, S929 (2006).

\bibitem{LiCs-dipole-moments} M.Aymar, O.Dulieu, J. Chem. Phys. {\bf 122}, 204302 (2005).

\bibitem{RbCs-DeMille} J.M.Sage, S.Sainis, T.Bergeman, D.DeMille, Phys. Rev. Lett. {\bf 94}, 203001 (2005).

\bibitem{Rb-opt-latt} T.Volz, N.Syassen, D.M.Bauer, E.Hansis, S.D\"urr, G.Rempe, Nature Phys. {\bf 2}, 692 (2006).

\bibitem{NaRb-hyperf} K.Matsubara, Y.-C.Wang, K.Ishikawa, M.Baba, A.J.McCaffery, H.Kato, J. Chem. Phys. {\bf 99}, 5036 (1993).

\bibitem{NaRb-hyperf1} S.Kasahara, T.Ebi, M.Tanimura, H.Ikoma, K.Matsubara, M.Baba, H.Kato, J. Chem. Phys. {\bf 105}, 1341 (1996).

\bibitem{NaK-hyperf} P.Burns, A.D.Wilkins, A.P.Hickman, J.Huennekens, J. Chem. Phys. {\bf 122}, 074306 (2005).

\bibitem{Raman-deexc} E.Taylor-Juarros, R.C$\hat{o}$t\'e, K.Kirby, Eur. Phys. J. D {\bf 31}, 231 (2004).

\bibitem{Feshbach} T.K\"ohler, K.G\'oral,P.S.Julienne, Rev. Mod. Phys. {\bf 78}, 1311 (2006).

\bibitem{BiN} O.Shestakov, E.H.Fink, Chem. Phys. Lett. {\bf 211}, 473 (1993); R.Breidohr, O.Shestakov, K.D.Setzer, E.H.Fink, J. Mol. Spectr. {\bf 172}, 369 (1995).

\bibitem{BaI-2Sigma-Hamilt} W.E.Ernst, J.K\"andler, C.Noda, J.S.McKillop, R.N.Zare, J. Chem. Phys. {\bf 85}, 3735 (1986).

\bibitem{SrI-hyperfine} W.E.Ernst, J.O.Schr\"oder, B.Schaal, Chem. Phys. Lett. {\bf 155}, 47 (1989).

\bibitem{alkali-halides} T.T\"orring, W.E.Ernst, J.K\"andler, J. Chem. Phys. {\bf 90}, 4927 (1989).

\bibitem{lattice-lifetime} G.Thalhammer, K.Winkler, F.Lang, S.Schmidt, R.Grimm, J.H.Denschlag, Phys. Rev. Lett. {\bf 96}, 050402 (2006).

\bibitem{lattice-switching} P.S.Julienne, C.J.Williams, Y.B.Band, M.Trippenbach, Phys. Rev. A {\bf 72}, 053615 (2005).

\end{thebibliography}
\end{document}